\documentclass[twocolumn]{aastex631}

\begin{document}

\title{Exploring the central region of NGC 1365 in the ultraviolet domain}

\author[0009-0002-8410-9937]{Kshama Sara Kurian}
\affiliation{Indian Institute of Astrophysics,
Koramangala, Block II, Bangalore 560034, India}
\affiliation{Pondicherry University,
Kalapet, Puducherry 605014, India}

\author[0000-0002-4998-1861]{C. S. Stalin}
\affiliation{Indian Institute of Astrophysics,
Koramangala, Block II, Bangalore 560034, India}
\email{stalin@iiap.res.in}

\author[0000-0003-2212-6045]{Dominika Wylezalek}
\affiliation{Astronomisches Rechen-Institut, Zentrum fur Astronomie der Universitat Heidelberg,
Monchhofstr. 12-14, 69120 Heidelberg, Germany}

\author[0000-0002-4381-0383]{Mariya Lyubenova}
\affiliation{European Southern Observatory, Karl-Schwarzschild-Str. 2, 85748 Garching bei München, Germany}

\author[0000-0003-4586-0744]{Tek Prasad Adhikari}
\affiliation{CAS Key Laboratory for Research in Galaxies and Cosmology, Department of Astronomy, University of Science and Technology of China, Hefei, Anhui 230026, China}
\affiliation{School of Astronomy and Space Science, University of Science and Technology of China, Hefei, Anhui 230026, China}

\author[0000-0001-5933-058X]{Ashish Devaraj}
\affiliation{Department of Physics and Electronics, CHRIST (Deemed to be University), Bangalore 560029, India}

\author[0000-0003-4973-4745]{Ram Sagar}
\affiliation{Indian Institute of Astrophysics,
Koramangala, Block II, Bangalore 560034, India}

\author[0000-0002-5908-1488]{Markus-Kissler Patig}
\affiliation{ESA - ESAC - European Space Agency, Camino Bajo del Castillo s/n, 28692 Villafranca del Castillo, Madrid, Spain}

\author[0000-0003-0793-6066]{Santanu Mondal}
\affiliation{Indian Institute of Astrophysics,
Koramangala, Block II, Bangalore 560034, India}



\begin{abstract}
Active galactic nuclei (AGN) feedback and its impact on their host galaxies are 
critical to our understanding of galaxy evolution. Here, we present a combined 
analysis of new high resolution ultraviolet (UV) data from the Ultraviolet 
Imaging Telescope (UVIT) on {\it AstroSat} and archival optical spectroscopic 
data from VLT/MUSE, for the Seyfert galaxy, NGC 1365. Concentrating on the 
central 5 kpc region, the UVIT images in the far and near UV show bright star 
forming knots in the circumnuclear ring as well as a faint central source. After 
correcting for extinction, we found the star formation rate (SFR) surface density 
of the circumnuclear 2 kpc ring to be similar to other starbursts, despite the 
presence of an AGN outflow, as seen in [OIII] 5007 \AA. On the other hand, we found fainter UV and thus lower SFR  in the direction south-east of the AGN 
relative to north-west in agreement
with observations at other wavelengths from JWST and ALMA. The AGN outflow 
velocity is  found to be lesser than the escape velocity, suggesting that the 
outflowing gas will rain back into the galaxy.  The deep UV data has also 
revealed diffuse UV emission in the direction of the AGN outflow. By 
combining [OIII] and UV data, we found the diffuse emission to be of AGN origin. 
\end{abstract}

\keywords{galaxies: Seyfert --- ultraviolet: galaxies --- galaxies: evolution}


\section{Introduction} \label{sec:intro}

Most galaxies in the Universe harbour supermassive black holes (SMBHs; 
10$^6-$10$^{10}$ M$_{\odot}$) at their centres \citep{1995ARA&A..33..581K,2013ARA&A..51..511K}.  These SMBHs while accreting 
matter from their surroundings lead to the trigger of the active galactic 
nuclei (AGN) as well as the growth of the host galaxy.  This close relationship 
between AGN and host galaxy is also evident in the
correlation observed between the mass of the SMBH (M$_{BH}$) in AGN and galaxy
luminosity \citep{1995ARA&A..33..581K}, M$_{BH}$ and galaxy mass
\citep{1998AJ....115.2285M} and M$_{BH}$ and the velocity
dispersion $\sigma$, of the stellar bulge (M-$\sigma$ relation; \citealt{2000ApJ...539L...9F}).
AGN thus play an important role in the evolution of galaxies via the injection of
energy into the gas in their host galaxies, a process called feedback
\citep{2012ARA&A..50..455F,2015ARA&A..53..115K}.

In many AGN, nuclear outflows from their central regions, in both ionized and molecular phases have been observed and these outflows can have a strong impact on the formation of stars in the central regions of AGN. They can either suppress star formation (negative feedback; \citealt{2012MNRAS.425L..66M,2017NatAs...1E.165H}) or enhance star formation (positive feedback; \citealt{2013MNRAS.433.3079Z,2020A&A...639L..13N}). Also, both positive and negative
feedback could operate in the
same system \citep{2019ApJ...881..147S,2021A&A...645A..21G}. A clear understanding 
of the impact AGN have on the central regions of their host galaxies requires high-resolution observations such as those from integral field spectroscopy (IFS) which can spatially resolve star formation (SF) and AGN outflows. Such high-resolution observations have become possible in recent years. However, combining such high spatial resolution observations, along with observations at other wavelengths such as sub-mm/UV and photoionization modelling would be a robust and better approach to establishing how the connection between SF and AGN works. At low redshifts, dominated by Seyfert-type AGN, we can spatially 
separate the AGN-dominated regions from the SF-dominated regions.  Here we report the results of one 
such study on a nearby AGN NGC 1365.

NGC 1365 is a low luminosity (L$_{bol}$ = 2 $\times$ 10$^{43}$ erg s$^{-1}$) 
Seyfert 1.8 type AGN at a distance of 18.6 Mpc \citep{2012MNRAS.425..311A,
1999A&ARv...9..221L}. At this distance, 1 arcsec corresponds to about 90 pc.  An 
early comprehensive review of the characteristics of the AGN in NGC 1365, can be 
found in \cite{1999A&ARv...9..221L}. It is found to show biconical ionised 
outflows in [OIII] \citep{1983MNRAS.203..759P, 2016MNRAS.459.4485L,2018A&A...619A..74V}. 
Star-forming clusters have also been found in the central regions of NGC 1365 
from infrared observations 
(\citealt{2012MNRAS.425..311A, 2019A&A...622A.128F,2023ApJ...944L..14W,2023ApJ...944L..19L}) 
and X-ray studies \citep{2009ApJ...694..718W}. This source has been extensively 
studied at high spatial resolution in the optical using IFS data from VLT/MUSE  
\citep{2018A&A...619A..74V} and optical spectra from TYPHOON \citep{2024ApJ...960...83S}, in the infrared using Herschel 
\citep{2021A&A...647A..86S}, in the near and mid-infrared
using JWST \citep{2023ApJ...944L..19L,2023ApJ...944L..14W,2023ApJ...944L..15S} 
as well as in the sub-mm using ALMA \citep{2021ApJ...913..139G,2023ApJ...944L..15S}. 
Although NGC 1365 is well studied in nearly all wavelength domains, the UV domain 
is unexplored. We aim to fill this gap and explore the relation between the AGN 
outflow and SF in the central 5 kpc square region (hereafter referred to as the 
'inner region') of the galaxy using spatially resolved optical IFS data and newly 
acquired UV data from the Ultra-Violet Imaging Telescope (UVIT) onboard 
{\it{AstroSat}} \citep{2006AdSpR..38.2989A,2022arXiv220304610S}, which is India's 
first space based multi-wavelength astronomical observatory.

\section{Observations}

\subsection{UVIT}

\begin{figure*}
\centering
\includegraphics[scale=0.9]{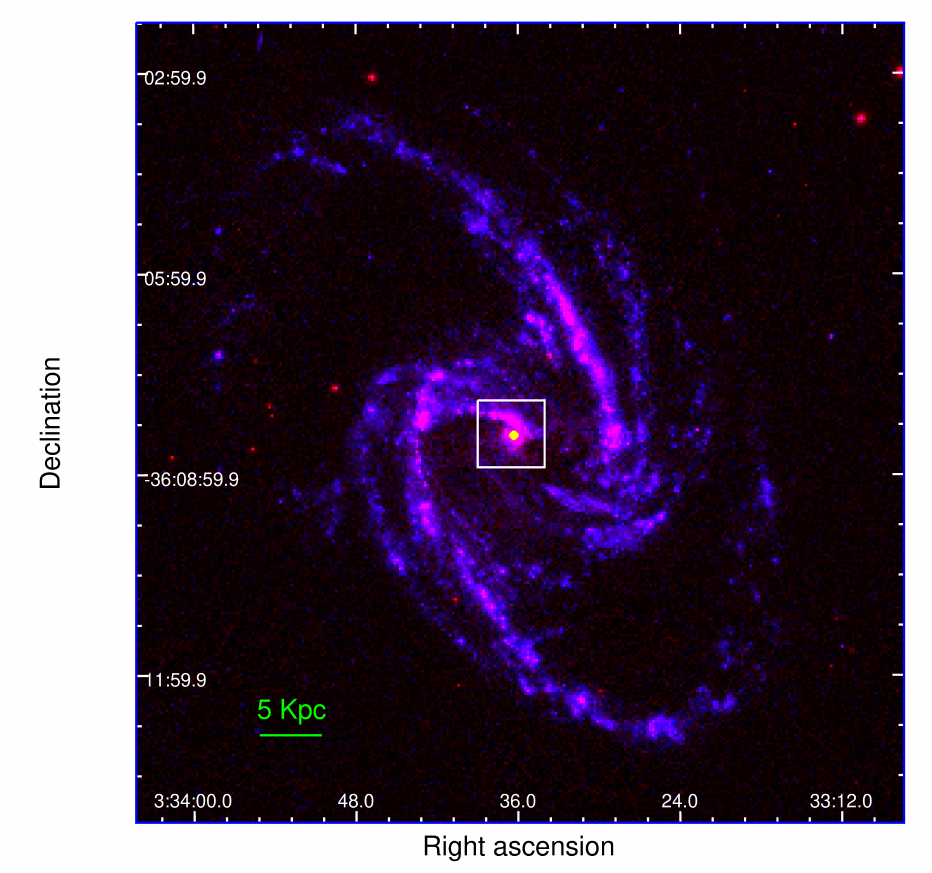}
\includegraphics[scale=0.5]{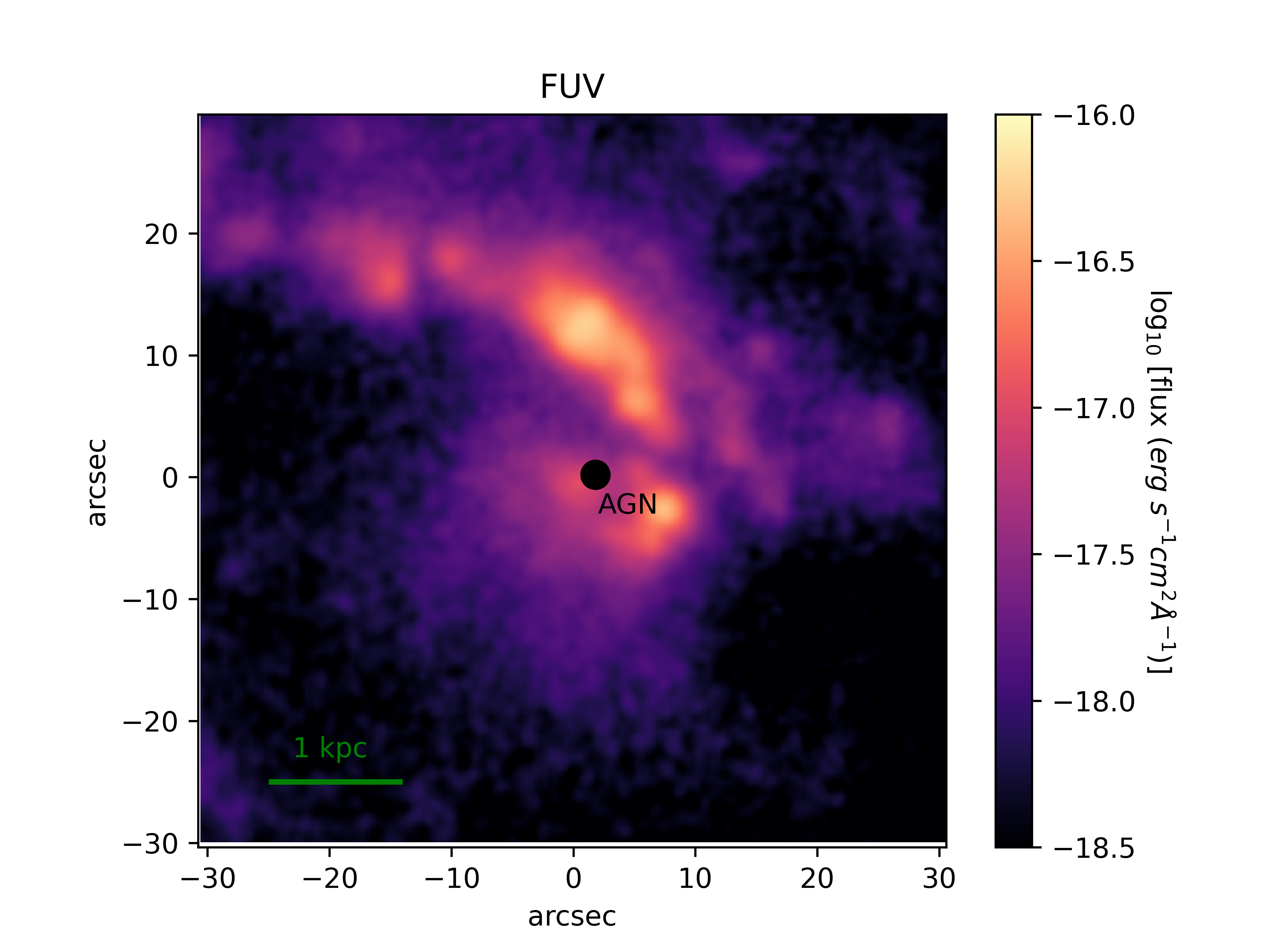}
\includegraphics[scale=0.5]{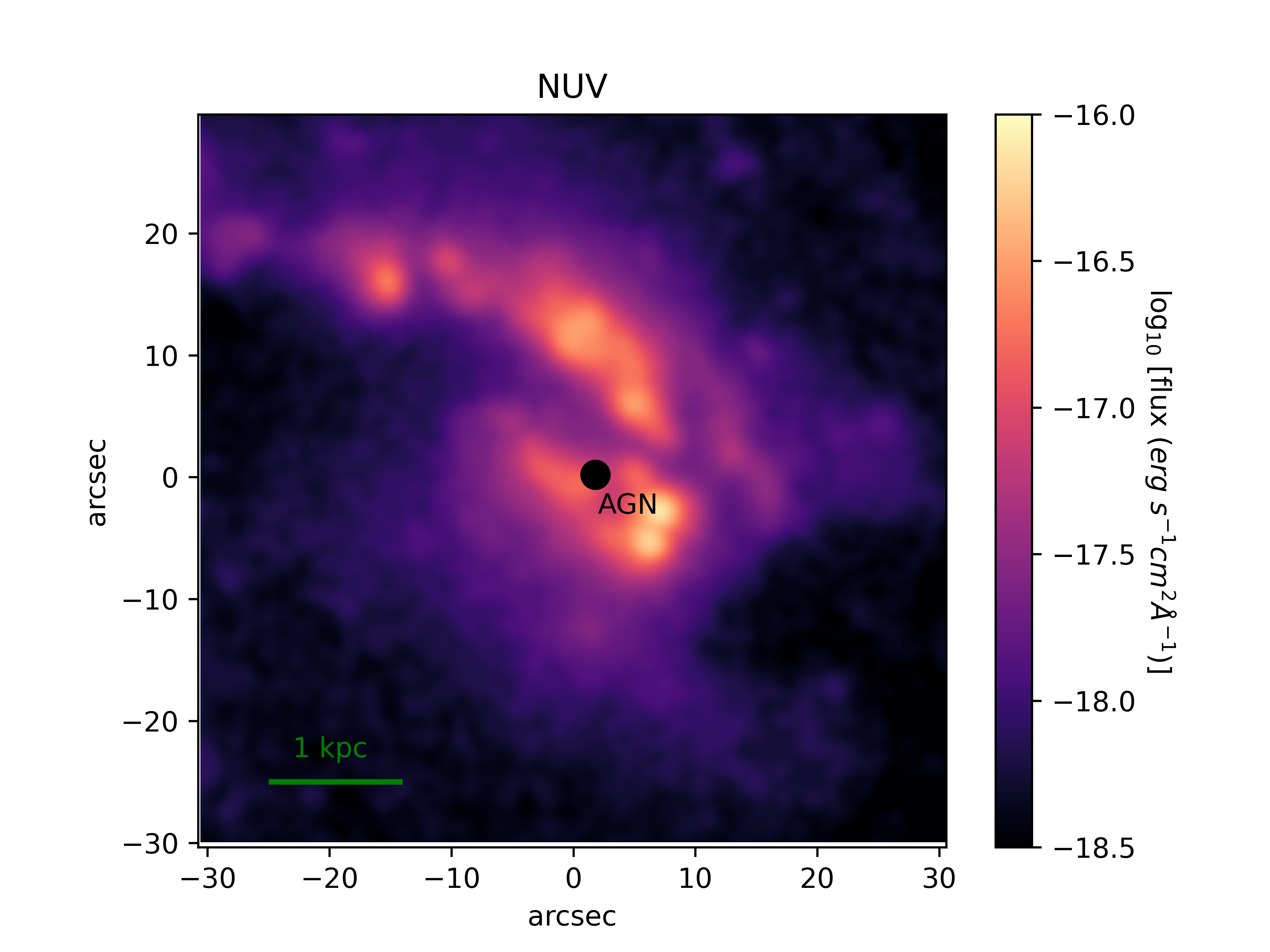}
\caption{Top panel: Composite UVIT image of NGC 1365 with FUV in blue 
and NUV in red.  The white box is the MUSE field of view 
($\sim 1^{\prime} \times 1^{\prime}$) corresponding to the central 5 kpc region {\bf which is the region analysed in this work}. The yellow dot in the centre of the box is the AGN position. Bottom 
panel: The UVIT FUV (left) and NUV (right) images of the inner {\bf 5 kpc} region of NGC 1365 {\bf corresponding to the MUSE field of view}. 
They have been corrected for Milky Way extinction. The black dot in the centre 
of the NUV and FUV images is the AGN position.}
\label{fig:figure-1}
\end{figure*}

NGC 1365 was observed using UVIT in two broad band filters
namely F169M (FUV; $\lambda_{mean}$ = 1608 \AA,  $\Delta\lambda$  = 
290 \AA) and N279M (NUV; $\lambda_{mean}$  = 2792 \AA,  $\Delta\lambda$ = 90 \AA) 
\citep{2017AJ....154..128T}. UVIT has a spatial resolution better than 1.5 arcsec 
and covers a field of view of $\sim$28 arcmin diameter.
The observations (PI: G. Dewangan) were done in the photon counting mode with a default
frame count rate of $\sim$29 frames per second \citep{2017AJ....154..128T}. We downloaded 
the science-ready level-2 (L2) images that correspond to the 
OBSIDs A02\_006T01\_9000000776, A02\_006T01\_9000000802 and A02\_006T01\_9000000934 
from the Indian Space Science Data 
Center (ISSDC), Bangalore. In the combined images from ISSDC, we found the exposure time to 
be lesser than the sum of the individual orbit-wise images. We, therefore, based 
our analysis on the reduced orbit-wise images. We first aligned the orbit-wise images using the Image Reduction and Analysis Facility (IRAF; \citealt{1986SPIE..627..733T}) and combined those 
aligned orbit-wise images to create the combined images filter-wise.
The net exposure times of the resulting images are 24905 seconds and 
37833 seconds in F169M and N279M filters respectively.  
Astrometry on the combined images was carried out using the astrometry.net 
package \citep{2010AJ....139.1782L}. The UVIT position of the AGN matches within 0.7 arcsec 
of the {\it Gaia} position. 
For NGC 1365 observations, the 
background in FUV  and NUV is $ ~ 1 \times 10^{-4} cps/arcsec^2 $ and 
$ 2.2 \times 10^{-4} cps/arcsec^2 $ respectively. 
The top panel of Fig. \ref{fig:figure-1} shows the composite 
image of NGC 1365 in FUV (blue) and NUV (red). Bright FUV and NUV knots of SF 
are seen in the inner region and the spiral arms. The bottom panel of 
Fig. \ref{fig:figure-1} shows the zoomed view of the FUV and NUV flux maps of 
the inner region that covers the MUSE field of view.

\subsection{MUSE}
The MUSE optical IFU observations of NGC 1365 were acquired on 12 October 2014 under the program 094.B-0321(A)  (PI: A. Marconi) \citealt{2018A&A...619A..74V}). For our data analysis, we used the fully reduced, 
4 ksec deep data cube of NGC 1365 which is available on the ESO archive. The 
MUSE field of view is 1$^{\prime}$ $\times$ 1$^{\prime}$ with a 0.2 arcsec/pixel 
spatial sampling. The median seeing during the observations 
was about 0.76 arcsec \citep{2018A&A...619A..74V}. The spectral window is 4750$-$9352 
\AA\; and the spectral binning is 1.25 \AA/channel.

\section{Data Analysis}

\subsection{MUSE}

\begin{figure*}
\includegraphics[scale=0.5]{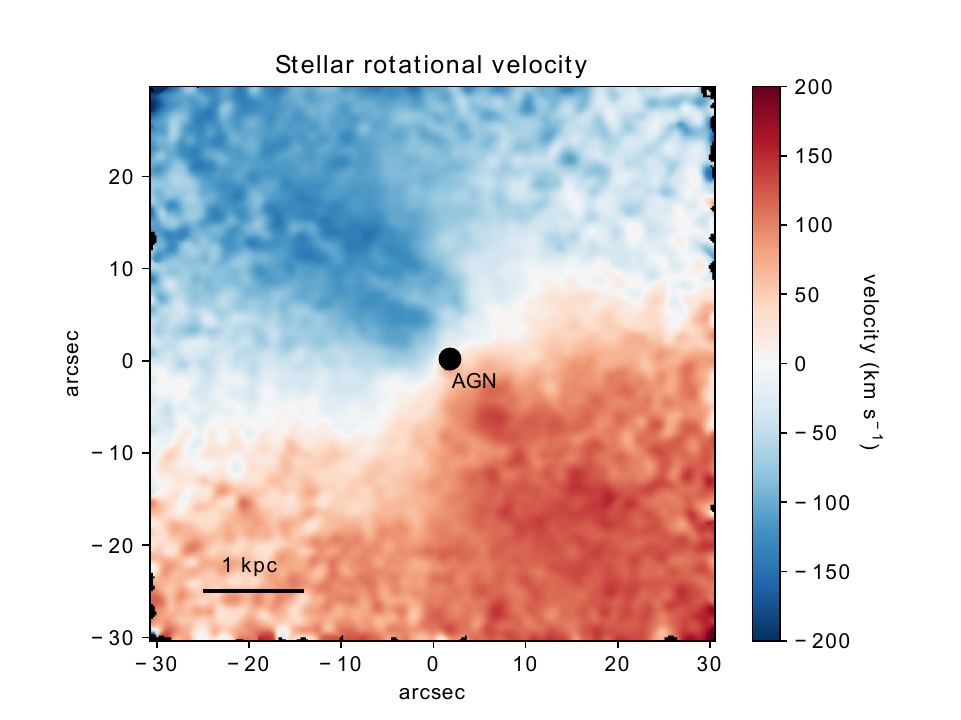}
\includegraphics[scale=0.5]{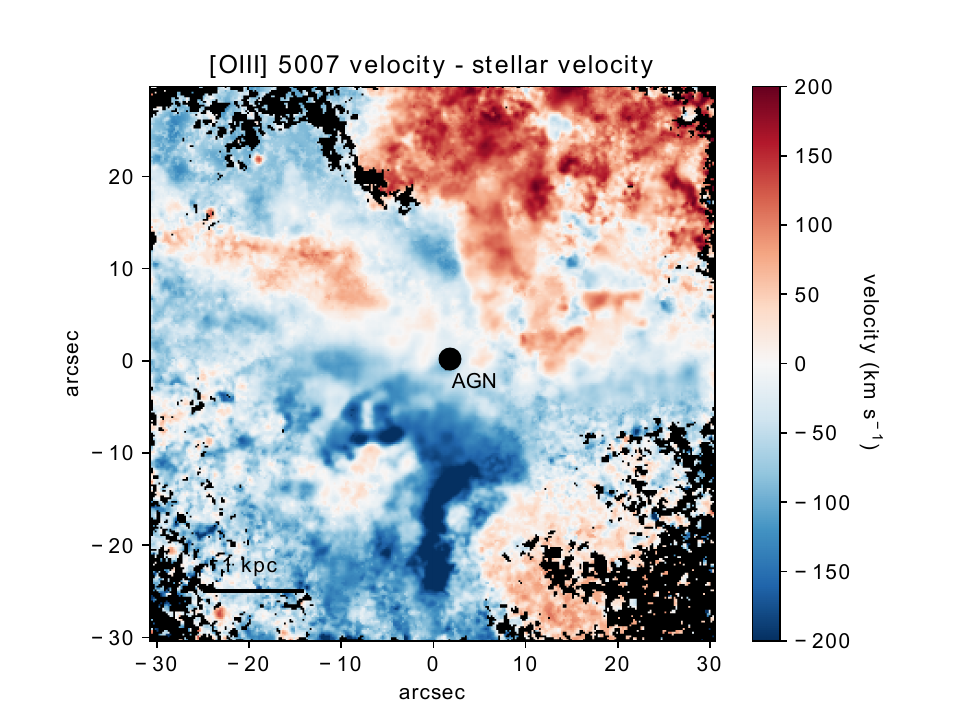}

\medskip

\includegraphics[scale=0.5]{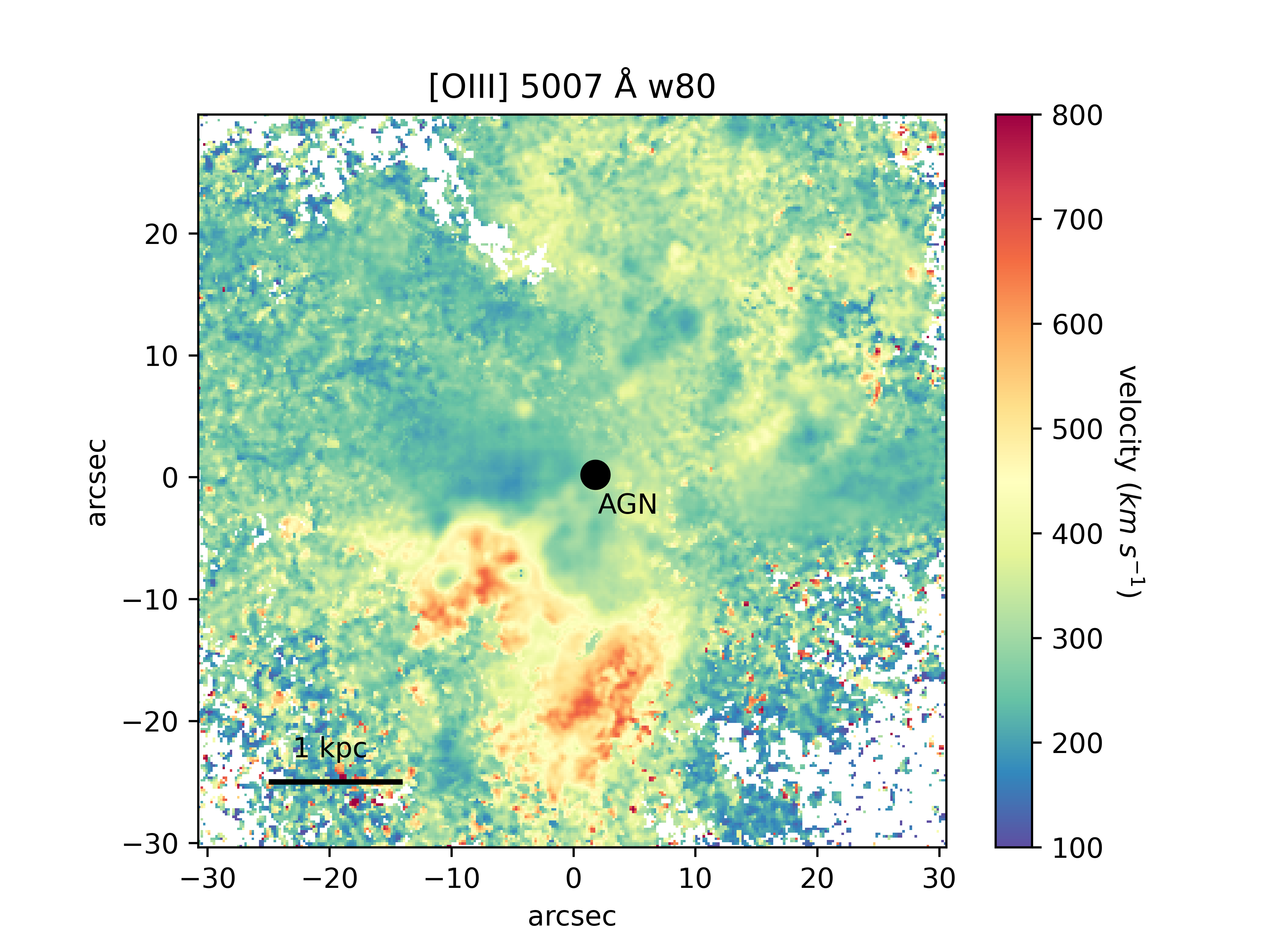}
\includegraphics[scale=0.5]{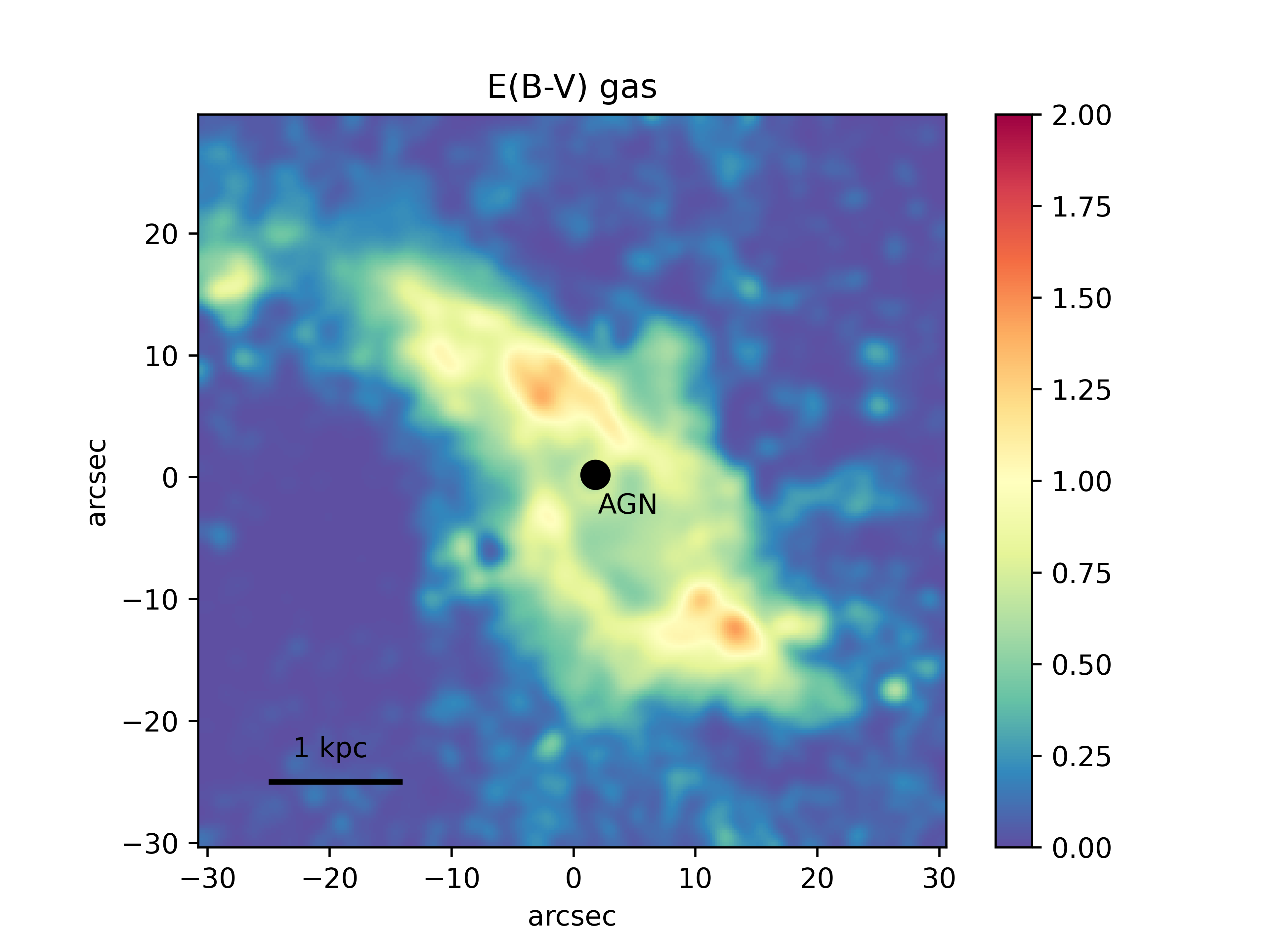}
\caption{The stellar rotational velocity map (top left), the stellar rotational 
velocity subtracted [OIII] {\bf 5007 \AA\ velocity} shift (top right), the W80 map of 
the [OIII] 5007 \AA\ line (bottom left) and the  E(B$-$V), of the nebular 
gas (bottom right). The AGN is marked based on the brightest H$\alpha$ region showing 
broad H$\alpha$ emission line width.}
\label{fig:figure-2}
\end{figure*}

We analysed the MUSE data using a combination of our own {\it python} 
scripts and the {\it python} package MPDAF (MUSE Python Data Analysis
Framework; \citealt{2016ascl.soft11003B}).  
The central 2 arcsec nuclear region was not considered for further 
analysis as we are not interested in the broad line region of NGC 1365.
To estimate the host galaxy kinematics, we fitted a single Gaussian 
to the Ca+Fe blend (MUSE
resolution allows the blend to be treated as a single line) at 
6495 \AA \ \citep{2009ApJS..183....1H}, after subtracting the stellar 
continuum. For this, spaxels with SNR (signal-to-noise ratio) $>$ 3 were considered. 
Here,  SNR is the ratio of the absolute value of the flux to the standard 
deviation of the line-free region of the continuum subtracted spectra. In the 
top left panel of Fig. \ref{fig:figure-2} 
we show the rotational velocity of the inner region of NGC 1365 as derived from 
the Ca+Fe blend. This is similar to that derived by \cite{2018A&A...619A..74V} 
using the Penalized PiXel Fitting (pPXF; \citealt{2004PASP..116..138C}) code.

We measure the flux of the [SII] doublet at $\lambda$6716, 
6731 \AA, H$\beta$-[OIII] line complex (H$\beta$ - $\lambda$4861\AA, [OIII] doublet - $\lambda$4959, 5007\AA) and
H$\alpha$-[NII] line complex (H$\alpha$ - $\lambda$6564.6 \AA, [NII] 
doublet - $\lambda$6549.8, 6585.2 \AA). To estimate
the flux of the emission lines, we fitted {\bf each individual} observed emission line with two  
Gaussian components (except for H$\beta$ and H$\alpha$ which required an additional 
Gaussian component to account for the absorption line) that reproduces the line 
profile well. During the fit, the 
intensity ratios of the [OIII] and [NII] doublets were fixed to 2.95 and
3.0 respectively \citep{2006agna.book.....O}. 
To separate shocked or Low-ionization nuclear emission-line (LINER) regions from 
AGN or Seyfert and starburst ionised gas, we used the [SII]-BPT (Baldwin, 
Phillips \& Terlivich; \citealt{1981PASP...93....5B,1987ApJS...63..295V}) 
diagram ([SII]6716,6731/H$\alpha$ vs [OIII]5007/H$\beta$). 
The BPT diagram indicates that the conical [OIII] 5007 \AA\; emission seen in 
the inner region of NGC 1365 is predominately due to photoionization from AGN 
and is confined to the north-west (NW) and 
south-east (SE) directions.

We determined the kinematics of the [OIII] emission line using the V50 and W80 
parameters \citep{2014MNRAS.442..784Z} from the total fitted line profile of the 
[OIII] 5007 \AA\; line. The parameter V50 is the velocity at 50$\%$ of the total 
flux and is a proxy to the velocity shift. W80 is the difference between the 
90$^{th}$ and 10$^{th}$ percentile of the total flux and is a proxy to the 
velocity dispersion (W80 = 1.088FWHM, \citealt{2016MNRAS.459.3144Z}). 
Fig. \ref{fig:figure-2} top right and bottom left panels show the kinematic 
maps derived from the [OIII] 5007 \AA\; line. The AGN photoionized 
cone is clearly seen to be outflowing in the stellar rotational velocity 
subtracted [OIII] 5007 \AA\; velocity shift map. The NW cone is red-shifted 
while the SE cone is blue-shifted. 
We found the AGN ionisation cone to have high W80 values of greater than 450 km/s. 

We estimated the colour excess E(B$-$V) as 
\begin{equation}
E(B-V) = 1.97 * log_{10}\left [\frac{{}(H_{\alpha}/H_{\beta})_{obs}}{{}(H_{\alpha}/H_{\beta})_{int}}  \right ]
\end{equation}
where (H$\alpha$/H$\beta$)$_{obs}$ and (H$\alpha$/H$\beta$)$_{int}$ is the observed 
and intrinsic H$\alpha$/H$\beta$ line ratio respectively.
This equation considers an intrinsic value for (H$\alpha$/H$\beta$) to be 
2.87 under Case B recombination conditions \citep{2006agna.book.....O}. 
Fig. \ref{fig:figure-2} bottom right panel shows the E(B$-$V) map of 
the nebular gas. We found the ring to have an average E(B$-$V) of $\sim$1 which 
translates to an A$_{v}$ of $\sim$3 (assuming R$_{v}$ = 3.1) and is
similar to that found by \cite{2018A&A...619A..74V}.

\subsection{UVIT}
The UVIT FUV and NUV images were converted to flux units 
using the unit conversion in \cite{2017AJ....154..128T}. In order to perform a combined study 
of the MUSE and UVIT data, the UVIT FUV and NUV images of NGC 1365 were cropped 
to the MUSE field of view. The position of the AGN in 
UVIT matches with the MUSE position within 0.4 arcsec.
 We then corrected the UV data for the Galactic 
extinction using \cite{1989ApJ...345..245C}, for a foreground 
extinction at the V band of 0.0543 mag, taken from \cite{2011ApJ...737..103S}. 
The intrinsic FUV and NUV fluxes were then estimated from the Milky Way corrected UV fluxes 
applying the \cite{2000ApJ...533..682C} reddening curve for the wavelength range, 1200 \AA \, $\leq \, \lambda \, \leq$\, 6300 \AA, which is
\begin{equation}
k_{\lambda} = 2.65\left(-2.15+ \frac{1.50}{\lambda}-\frac{0.19}{\lambda^{2}}+\frac{0.01}{\lambda^{3}}+R_{v}\right) 
\end{equation}
Here, R$_{\textup{v}}$ = 3.1. Following \cite{2000ApJ...533..682C},  
we estimated the colour excess of the stellar emission from the colour excess of 
the nebular emission using, $E_{s}(B-V) = 0.44E(B-V)$.
We then calculated the intrinsic UV flux using the equation, 
\begin{equation}
F_{int}(\lambda)=F_{obs}(\lambda)10^{0.4E_{s}(B-V)k_{\lambda}}
\end{equation}

\section{Results}

\subsection{\textbf{UV slope-$\beta$}}

\begin{figure}
\includegraphics[scale=0.5]{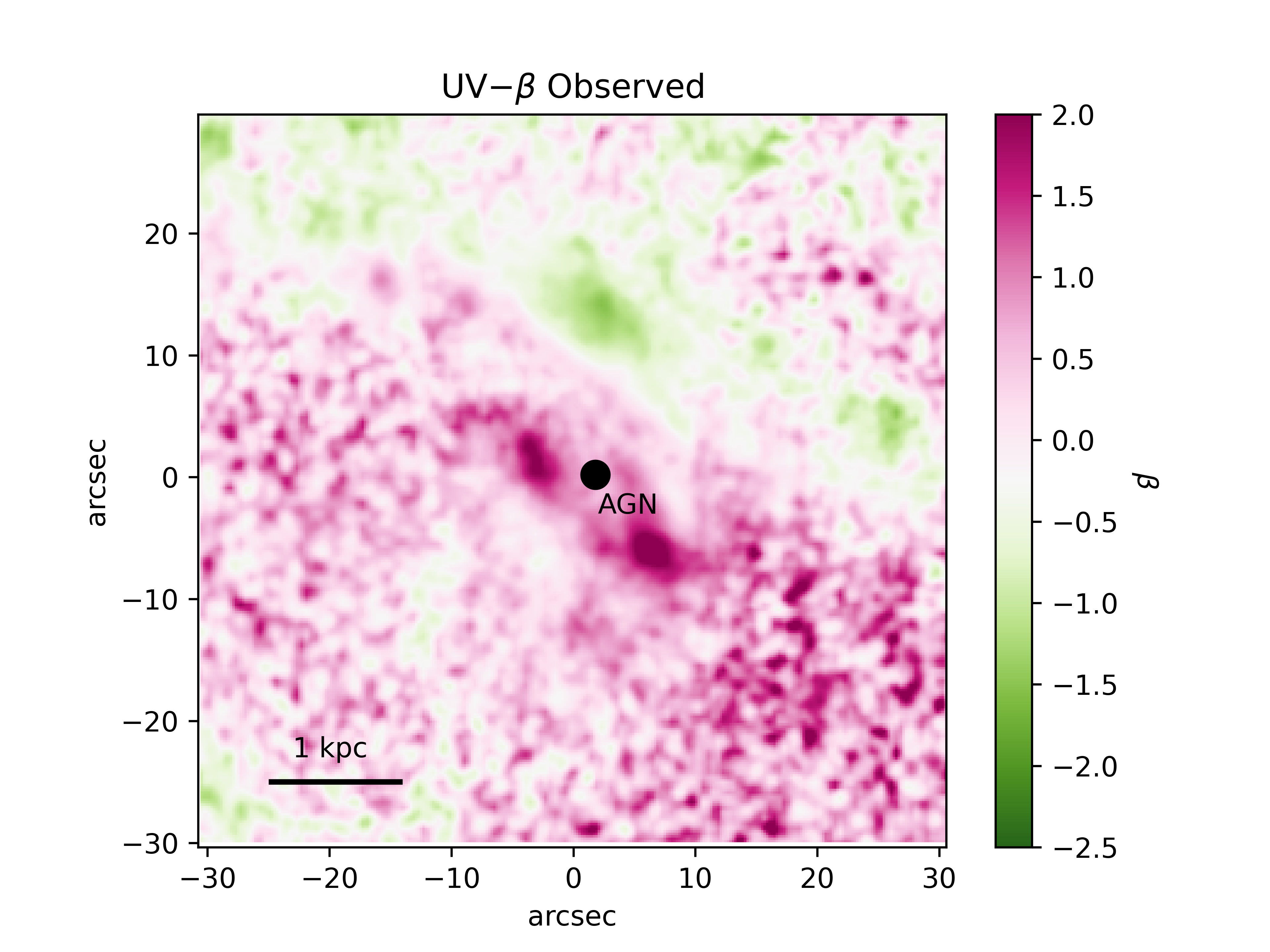}
\includegraphics[scale=0.5]{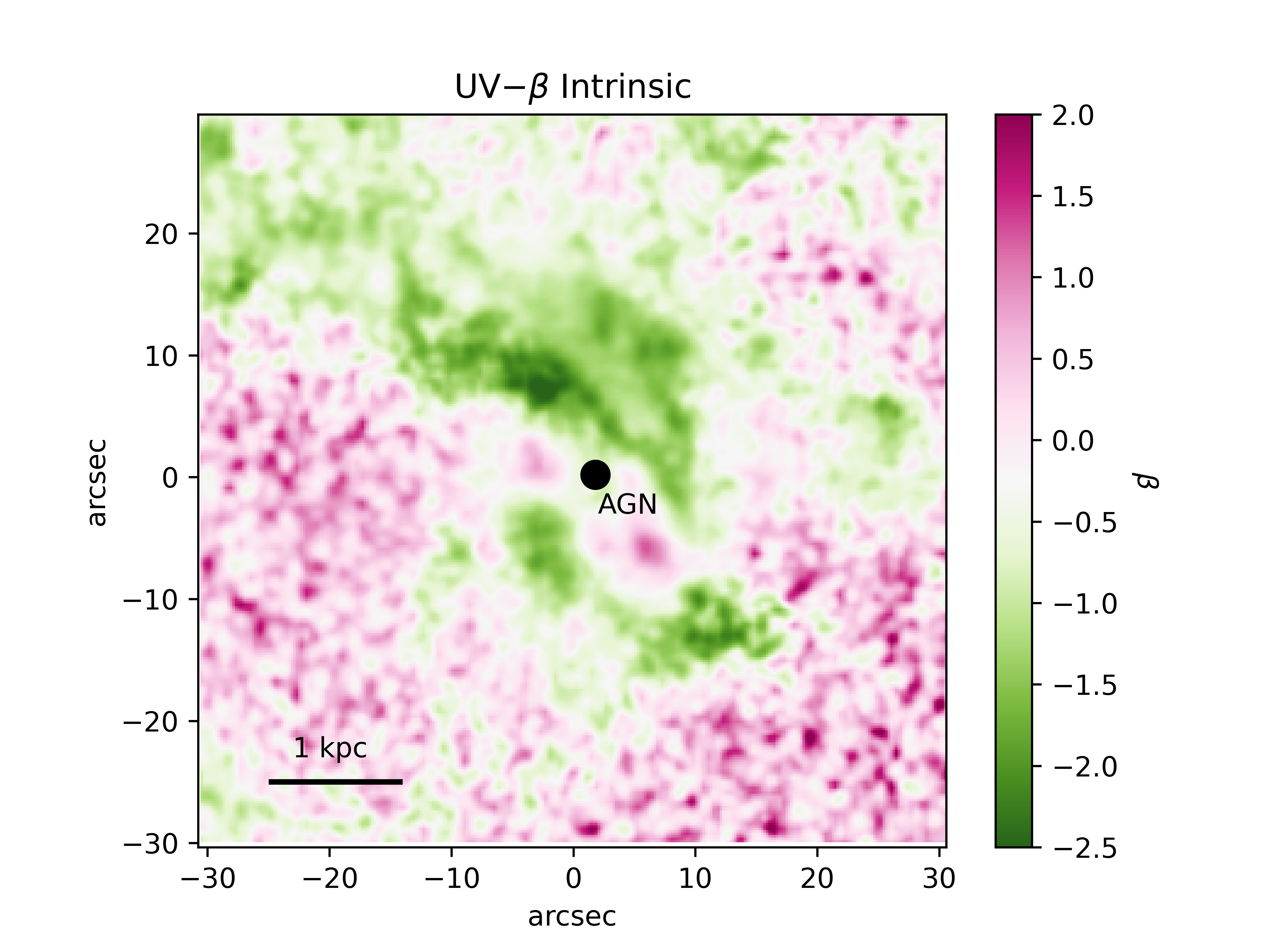}
\caption{Map of the observed (top) and intrinsic (extinction corrected, bottom) 
UV slope, $\beta$ of the inner region of NGC 1365. The extinction corrected 
$\beta$ map shows negative $\beta$ values (green coloured) in the regions 
predominantly occupied by {\bf massive star formation} as seen in the H$\alpha$ map in \cite{2018A&A...619A..74V}.}
\label{fig:figure-3}
\end{figure}

The UV continuum slope is a proxy to the UV colour from which age and 
other stellar population characteristics can be derived  
\citep{2001PASP..113.1449C}. The UV slope\, $\beta$, is defined as the slope of the power 
law function, F$_{\lambda} \; \propto \; \lambda^{\beta}$ where F$_{\lambda}$ is 
the flux density. We created a map of the observed and intrinsic (extinction 
corrected) UV slope, $\beta$ which is shown in Fig. \ref{fig:figure-3}. 

The intrinsic $\beta$ map shows the star-forming regions in green with 
negative $\beta$ values. The 
deepest green regions have $\beta$ values close to $-$2.5 which is the dust 
free $\beta$ value of a 5 Myr stellar population assuming an instantaneous burst of 
SF and solar abundance (refer to Table 6 in 
\citealt{2001PASP..113.1449C}). 

\subsection{\textbf{Star formation rate (SFR)}}

\begin{table}[ht!]
\begin{center}
\caption{SFR of the bright star-forming ring in the central region of diameter 2.2 kpc of NGC1365 (r = 12$^{\prime\prime}$, $1^{\prime\prime}$ $\sim$ 90 pc, Kroupa IMF ).}
\label{table:parameters}
\begin{tabular}{lccccc}

\hline
Attenuation law & SFR FUV  &  SFR NUV &  SFR H$\alpha$\\

 &  M$_{\odot}$yr$^{-1}$ &  M$_{\odot}$yr$^{-1}$ & M$_{\odot}$yr$^{-1}$
 \\\hline
Uncorrected                        &   0.22  & 0.82 & 0.60 \\
Starburst, R$_{\textup{v}}$ = 3.1  &   2.68  & 4.76 & 3.12 \\
Starburst, R$_{\textup{v}}$ = 4.05 &   3.66  & 6.34 & 6.41 \\
Milky Way                          &   1.90  & 4.40 & 3.51 \\
SMC                                &   7.37  & 4.31 & 2.67 \\
\hline
\end{tabular}
\end{center}
\end{table} 

There are several SFR estimates in the literature, for the central regions of 
NGC 1365 \citep{2021ApJ...913..139G,2019A&A...622A.128F,2012MNRAS.425..311A}. They mainly use H$\alpha$ flux values which is an indirect estimator 
of SFR. On the contrary, UV flux directly traces the radiation from young star-forming regions and thus is the most appropriate estimator of SFR. Using the 
high-resolution extinction corrected UVIT FUV and NUV data, we derived new 
estimates of SFR. Most of the SF is said to be taking place in the Inner Lindblad 
region \citep{2005A&A...438..803G}, mainly concentrated in a star-forming ring 
of approximate diameter 2 kpc \citep{2012MNRAS.425..311A}. We estimated the 
intrinsic FUV and NUV SFR for the central 12 arcsec radius region after masking the central 2 arcsec radius region which is dominated by AGN light (the 12 arcsec region is shown
as a circle in Fig.\ref{fig:figure-44}, left panel). We estimated the SFR using the following equation from \cite{2013seg..book..419C}, assuming a Kroupa IMF: 
\begin{equation}  
SFR = 3 \times 10^{-47} \lambda L_{\lambda} \; (M_{\odot}yr^{-1})\\
\end{equation}
The UV SFRs are highly dependent on extinction correction, and so, to be 
thorough, we estimated the SFR for \cite{2000ApJ...533..682C} reddening curve, 
with a starburst R$_{\textup{v}}$ = 4.05 and 3.1, the
Small Magellanic Cloud (SMC) attenuation curve \citep{2003ApJ...594..279G} and 
the Milky Way reddening curve \citep{1989ApJ...345..245C}. For comparison, we also 
estimated the SFR from H$\alpha$ flux values. Both the UV and H$\alpha$ SFRs 
are shown in Table \ref{table:parameters}. \cite{2012MNRAS.425..311A} estimated 
an SFR of 5.6 M$_{\odot}$yr$^{-1}$ from H$\alpha$ and Spitzer 24 micron IR flux, 
for a Salpeter IMF.  We also 
derived the SFR surface density of NGC 1365 for a radius of 900 pc. It is found to be between 0.76 and 
2.94 M$_{\odot}$/yr/kpc$^{2}$ for SFR estimated from FUV and between 1.72 and 
2.53 M$_{\odot}$/yr/kpc$^{2}$ for SFR estimated from NUV, assuming the various 
attenuation curves. This is well within the expected SFR surface density at 
1 kpc scale from the centre of the galaxy for other starbursts 
\citep{1998ApJ...498..541K, 2012A&A...544A.129V}.

\subsection{\textbf{Spatial variation in SFR}}
\begin{figure}
\includegraphics[scale=0.55]{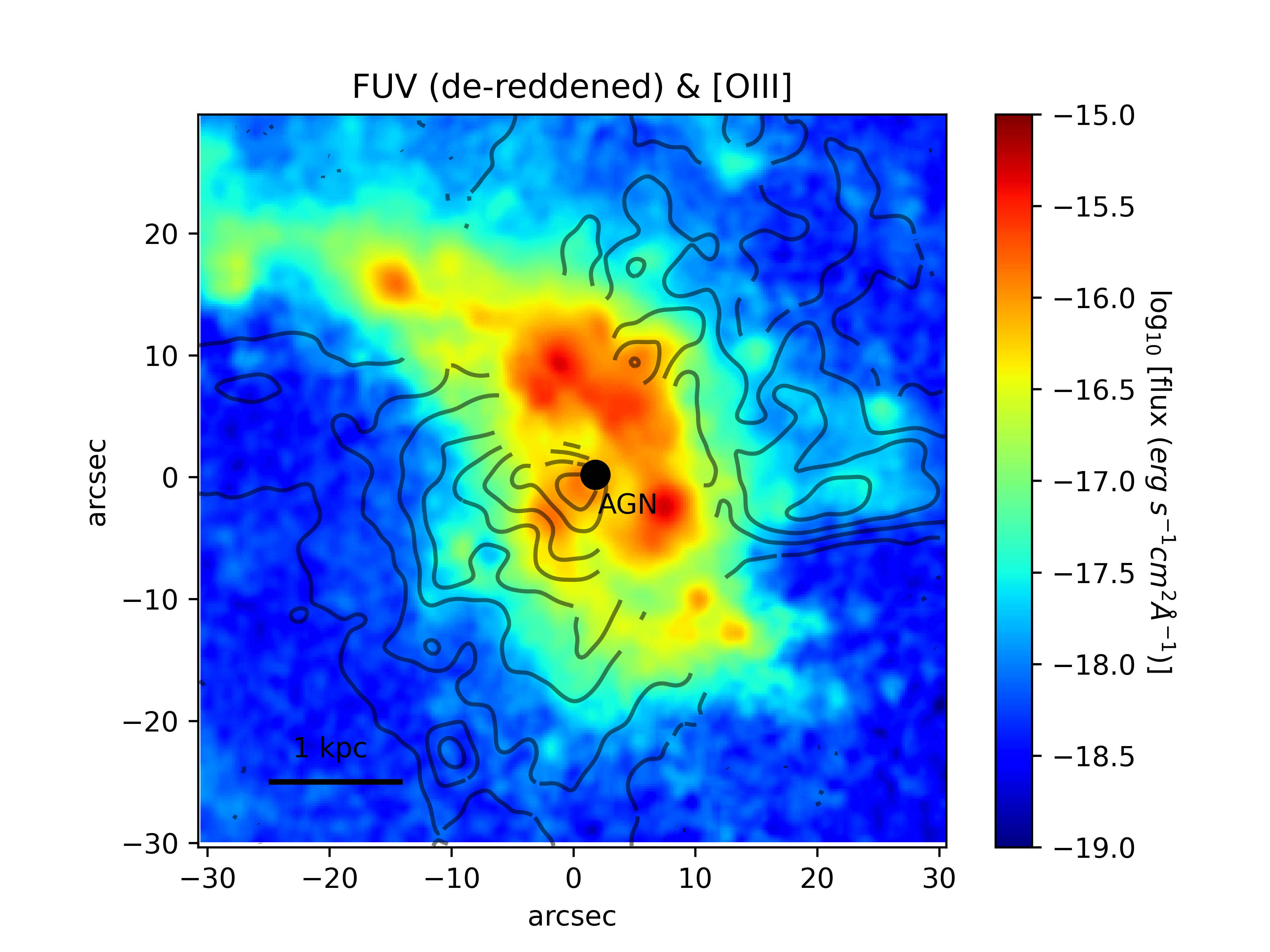}
\caption{Extinction corrected, intrinsic UVIT FUV flux map of the inner region of 
NGC 1365, with MUSE [OIII] 5007 \AA\; contours overlaid in black. Only the 
AGN-dominated regions are shown in the [OIII] contours after masking the LINER 
and starburst-dominated regions using the [SII]-BPT diagram.}
\label{fig:figure-5}
\end{figure}

In Fig. \ref{fig:figure-5}, we show the extinction corrected (following 
\cite{2000ApJ...533..682C} with R$_{\textup{v}}$ = 3.1) intrinsic FUV and 
NUV SFR map of the inner region with the [OIII] 5007 \AA\; AGN outflow contours 
overlaid in black. The starburst and LINER-dominated regions have been masked 
in the [OIII] outflow contours using the [SII]-BPT diagram and hence the [OIII] 
contours in this image show only the regions that are AGN-dominated. 

The intrinsic UV flux maps are directly proportional to the spatially resolved 
SFR in the inner region. 
In NGC 1365, we found a spatial variation of SFR in the star-forming ring.  
Specifically, the SE part of the ring has low FUV (lower SFR) with 
respect to the NW side of the ring. The low SFR regions are co-spatial with the 
regions showing strong [OIII] contours in the SE cone possibly hinting at the outflow affecting the SF. 
However, star forming rings seen in galaxies are generally made of several
aligned knots of SF and the SF can also be patchy 
\citep{2010MNRAS.402.2462C,2014A&A...562A.121C}. Recently using high resolution 
observations of the inner region 
of NGC 1365 from ALMA and JWST coupled with simulations, \cite{2023ApJ...944L..15S} 
found that the distribution of SF in its
central region is in-homogeneous. The northern side of the AGN has high SF, while
the southern side has reduced SF, though the molecular gas is similar on both the
northern and southern sides. According to them, the SF in the central region is driven by
gas inflow via bar and SF in the southern regions has not yet started. They conclude that the AGN outflow
has thus no impact on the gaseous disk. Considering the inclination {\bf of the AGN outflow and the galactic disk}, the SE cone is above the disk with its axis 5\textdegree \ within the galaxy axis \citep{1996A&A...305..727H}. Hence it seems less likely to intersect the galaxy plane. From studies of molecular gas kinematics, 
\cite{2023ApJ...944L..19L} too do not find evidence of the ionized outflow to intersect the
molecular gas disk. Recent high resolution observations are thus not in agreement
with the scenario proposed by \cite{2021ApJ...913..139G}  wherein the outflow 
impacts the molecular disk and causes AGN feedback, based on poor resolution 
ALMA maps. From UVIT observations, we found fainter UV and thus lower SFR in the
SE side relative to the NW side of the AGN which is in agreement with the recent
high resolution observations from JWST and ALMA 
\citep{2023ApJ...944L..15S,2023ApJ...944L..19L,2023ApJ...944L..14W}.

\subsection{\textbf{Diffuse UV light observed in the outflow cone}} 

\begin{figure*}
\centering
\includegraphics[scale=0.5]{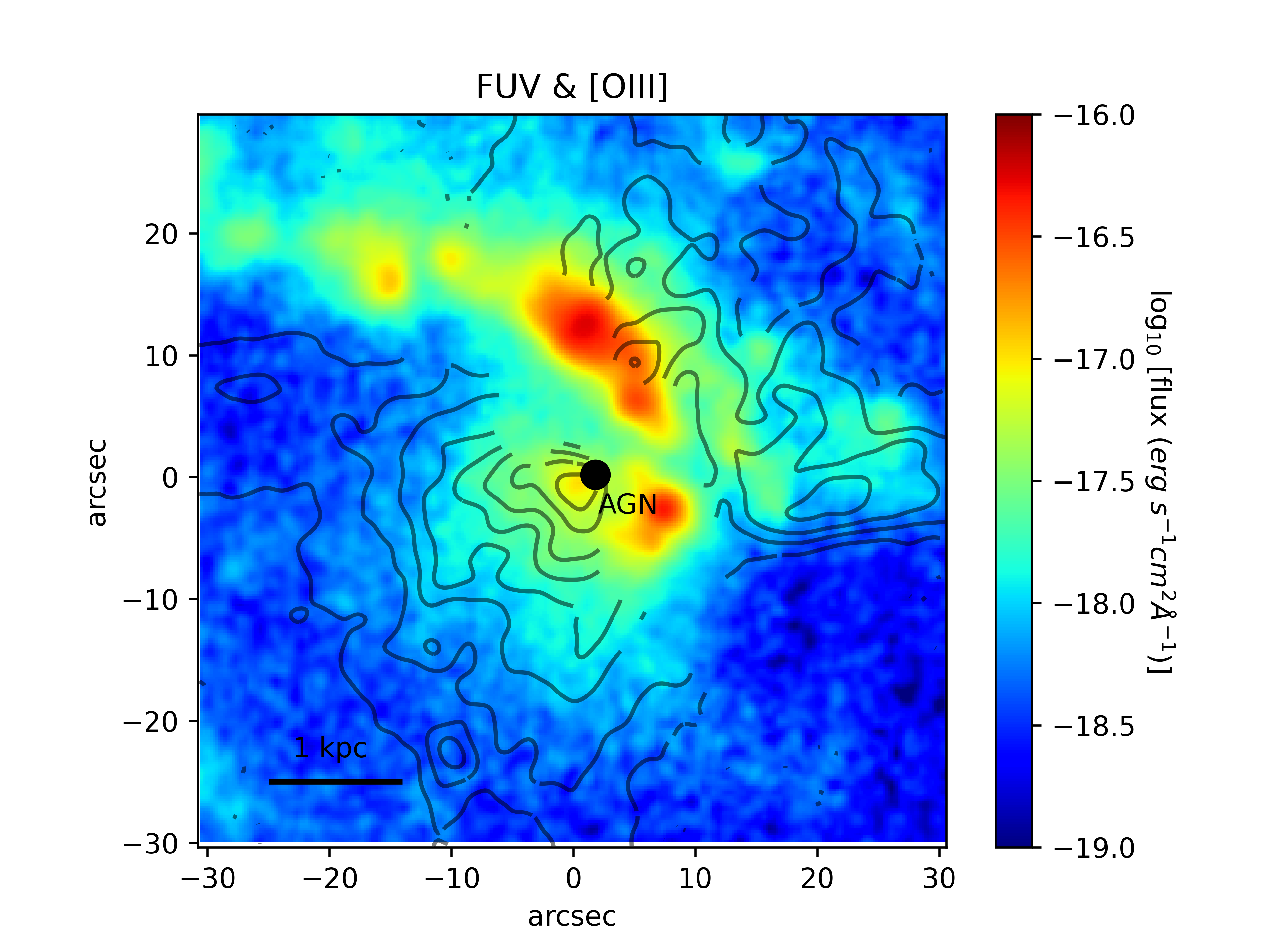}
\includegraphics[scale=0.5]{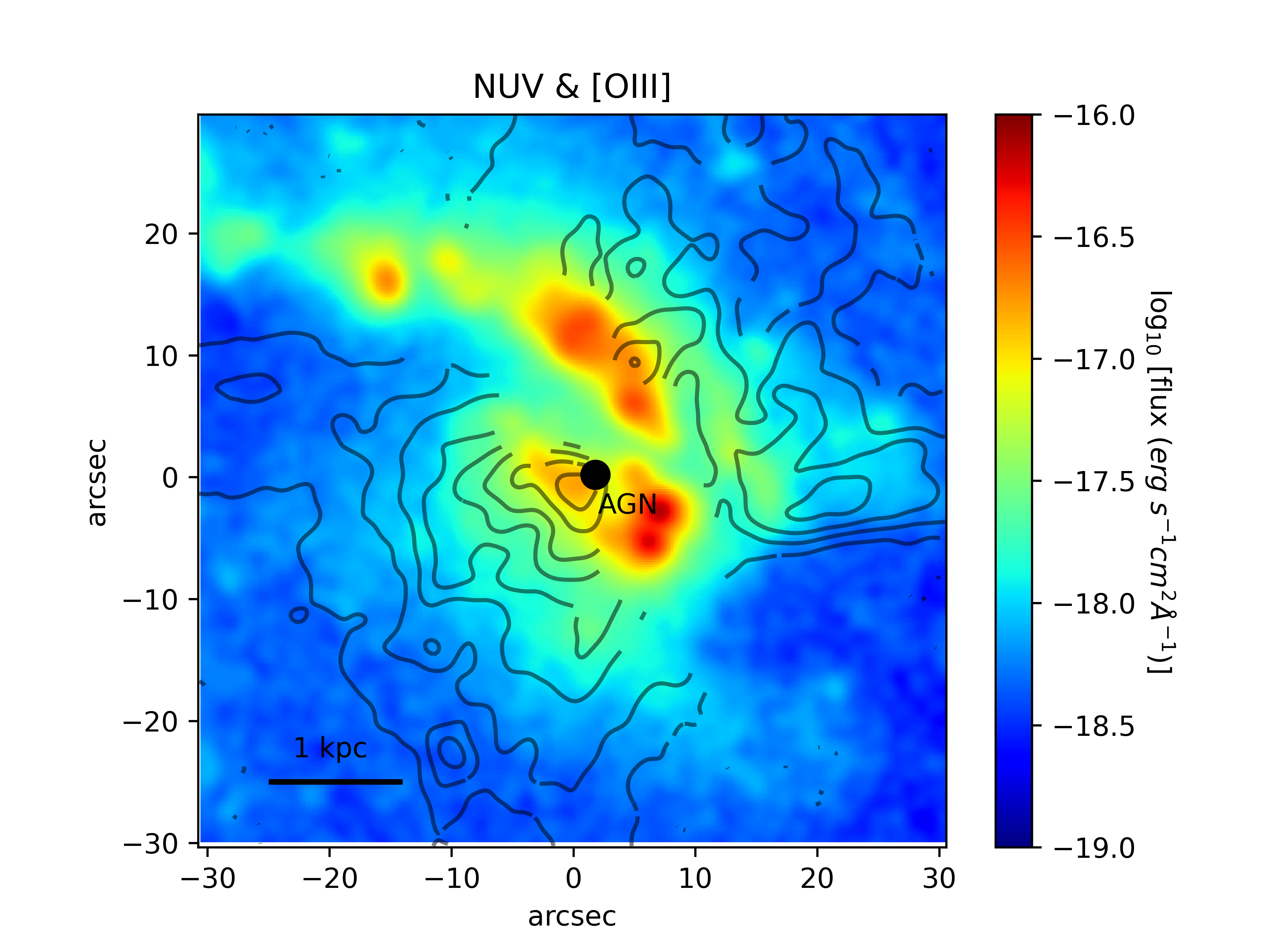}

\medskip

\includegraphics[scale=0.5]{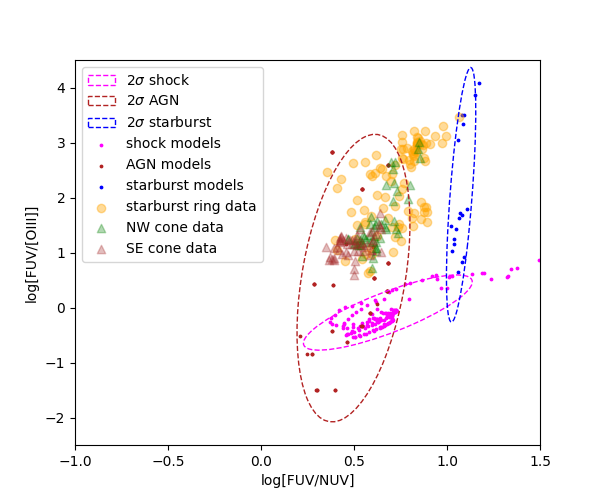}
\includegraphics[scale=0.5]{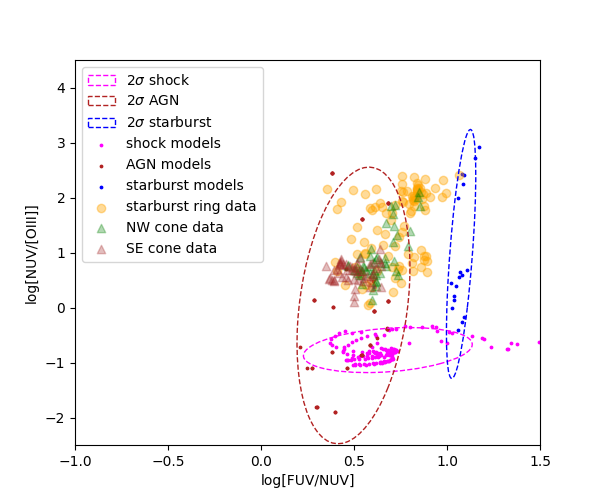}
\caption{Results from the combined UVIT and MUSE observations: Top panel: Galactic 
extinction corrected FUV (left) and NUV (right) maps of the inner region of 
NGC 1365, with AGN dominated [OIII] 5007 \AA\; contours from MUSE observations 
overlaid in black. Diffuse FUV and NUV emission is seen co-spatial to the [OIII] 
cone both in the NW and SE directions. Bottom panel: Flux ratios, FUV/[OIII] 
versus FUV/NUV  (left) and NUV/[OIII] versus FUV/NUV (right) of the starburst 
(yellow circles), AGN NW cone (green triangles) and AGN SE cone (red triangles) 
in the inner region of NGC 1365. Model flux ratios of AGN and starburst 
photoionized gas are shown in red and blue points and shocked gas is shown in 
magenta. Confidence ellipses having 95\% confidence are drawn around their 
respective model values.}
\label{fig:figure-4}
\end{figure*}

The UV morphology of the central regions of NGC 1365 predominantly points to a starburst origin of UV light when compared with the H$\alpha$ flux map. Along with the bright starburst regions, there is diffuse UV light in the NW and SE side of the AGN-dominated cone which is, in parts, similar to the morphology seen in the [OIII] outflow. This is shown in the Galactic extinction corrected UV images in the top panel of Fig. \ref{fig:figure-4}.

To determine the origin of this diffuse UV light, which is co-spatial with the 
[OIII] outflow region, we compared UV and [OIII] flux values 
to shock and photoionization models. 
We first calculated the flux ratios of FUV/[OIII], NUV/[OIII] and FUV/NUV for 
the outflow and the starburst ring of NGC 1365 after smoothing the [OIII] line 
emission map (by convolving with a Gaussian) to match the spatial resolution of 
UVIT data. To estimate the flux 
ratios, we used the \cite{2000ApJ...533..682C} (R$_{\textup{v}}$ = 3.1) 
extinction corrected FUV and NUV fluxes multiplied with their respective 
bandwidths ($\Delta\lambda$ = 290 \AA~ for FUV and $\Delta\lambda$ = 90 \AA~ for NUV). The starburst region was taken to be the same as the area used to 
determine the SFR in Section 4.2. This region was divided into smaller regions 
and the mean flux ratio in each section was calculated. The outflow region was 
roughly ascertained to be a double cone centred at the AGN, having a full opening 
angle of 90\textdegree, by using the [SII]-BPT AGN mask. This value of the 
opening angle is within the maximum opening angle of 100\textdegree\, modelled 
by \cite{1996A&A...305..727H}. 

We then estimated AGN and starburst photoionization as well as shock model flux ratios. The model values of the gas photoionized by AGN and starburst were generated using CLOUDY 17.0 \citep{2017RMxAA..53..385F}. 
For the AGN photoionization model, we considered a standard AGN continuum, with a big blue bump (BBB) temperature of 10$^5$K, an X-ray to UV slope of $\alpha_{ox} = -1.5$, a low-energy BBB slope of $\alpha_{uv} = -0.5$, and an X-ray power-law index of $\alpha_{x} = -1.35$ \citep{2009A&A...505..589G}. We computed the AGN models under isobaric conditions assuming plane parallel geometry \citep{2018ApJ...856...78A}, with an initial pressure of $P \;= \; 2 \times 10^7$ K \citep{2018MNRAS.479.4907D}. We adopted Milky Way dust grain abundance for solar metallicity. We generated a grid of spectra for photoionized gas, spanning a large range in ionisation parameters: log(U) = [-3.5, 2.0], hydrogen density log($n_H$) = [3, 6] (these are typical densities of the AGN narrow line region) and total column density log($N_H$) = [19, 24]. 
For the starburst models, we generated spectra of spherical HII regions 
photoionized by instantaneous starburst continuum whose spectra with stellar
masses that span the range between 1 M$_{\odot}$ and 100 M$_{\odot}$  were taken 
from Starburst99 \citep{1999ApJS..123....3L} for ionisation parameter: 
log(U) = [-4.0, -2.0], hydrogen density log($n_H$) = [1, 3] and starburst ages 
of 1 and 5 Myr. Similar to the AGN models, we adopted Milky Way dust grain 
abundance for solar metallicity for the starburst models. 

The shock models were generated from MAPPINGS III \citep{2008ApJS..178...20A} 
for shock velocities of 100 to 1000 km/s assuming pre-shock gas densities of 
0.1$-$1 cm$^{-3}$ and magnetic field values of 1$-$10 $\mu$G.  We then compared 
the extinction corrected flux ratios (applying reddening curve by 
\cite{2000ApJ...533..682C}, for an R$_{\textup{v}}$ = 3.1) to AGN
photoionization, starburst photoionization and shock model predictions. Fig. \ref{fig:figure-4} bottom panel shows the flux ratio map of FUV/[OIII] versus FUV/NUV and NUV/[OIII] versus FUV/NUV. 

The flux ratios, FUV/[OIII] and NUV/[OIII], of the outflow region along with their FUV/NUV values are found to match with AGN models. This indicates that the diffuse UV emission in the outflow region is predominantly of AGN origin. This could be due to line emission from AGN photoionization of gas clouds in the outflow \citep{2016MNRAS.456.3354F} as well as direct and scattered AGN continuum light from electrons or dust in the outflow \citep{2005AJ....129.1212Z, 2016MNRAS.456.2861O}.  

Geometrically, the outflow cone is said to have an opening angle of 100\textdegree\,\citep{1996A&A...305..727H} or a half opening angle of 50\textdegree\, with an inclination from the line of sight of 40\textdegree \citep{1995AJ....110.2037J}. From modelling the torus, \cite{2012MNRAS.425..311A} estimated an opening angle of 36\textdegree. Since the torus is thought to confine the outflow opening angle, this value suggests a narrower cone. If we assume the outflow cone half opening angle of the SE cone is 36\textdegree\, and is inclined at 40\textdegree\, from the line of sight, this would limit the contribution from the direct AGN continuum. However, electrons or dust in the cone can scatter the AGN continuum. Alternatively, there can also be a contribution from two-photon decay and hydrogen 
fluorescence to diffuse UV emission \citep{2022PASP..134h4302K}. The fact that the diffuse emission is detected in both FUV and NUV, as well as in both the southern and part of the northern cones, may provide the strongest support to the scattered light origin.  This is also in accord with the results found by \cite{2005AJ....129.1212Z} and \cite{2016MNRAS.456.2861O} in scattering cones of AGN.

Our observations indicate that deep UV data can also be used to reveal 
scattered light by dust in the AGN outflow. To the best of our knowledge, such 
a study, combining photometric UV data with spectroscopic [OIII] emission, has 
not been undertaken in Seyfert galaxies and this is the first study of its kind.

\newpage

\subsection{\textbf{Impact of AGN feedback in NGC 1365}}

From the estimation of the UV SFR surface density, the AGN outflow in NGC 1365 
seems to have no impact on the total SFR in the circumnuclear ring as the SFR 
surface density is comparable to other starbursts. Also, the SFR variation
seen in the central ring need not be due to the observed AGN outflow. NGC 1365 is 
a barred galaxy and in such galaxies, gas flows into the nuclear rings via the 
dust lanes. 
From hydrodynamical simulations, it is found that variations in the gas inflow rate can lead
to changes in SF in the nuclear ring, including lopsided star formation \citep{2023MNRAS.523.2918S}. 
Therefore inflow has a causal connection in regulating SF in nuclear rings 
\citep{2020MNRAS.497.5024S,2021ApJ...914....9M,2022ApJ...925...99M}. Lopsided
star formation in nuclear rings is known from observations \citep{2021MNRAS.505.4310C}.
Recent high resolution
observations of NGC 1365, supported by simulations, also favour a scenario 
wherein the varied SFR in the ring is due to gas inflow \citep{2023ApJ...944L..15S}, 
We thus conclude that the ionized AGN
outflow has no impact on the molecular gas in the central region of NGC 1365
and thus the observed lopsided star formation in the ring on either side of the AGN 
in NGC 1365 is explainable without invoking AGN feedback \citep{2023ApJ...944L..15S}.


The ionised gas in the [OIII] outflow has velocities of the 
order of 100$-$150 km/s. (\citealt{2018A&A...619A..74V}; this work). 
We estimated the escape velocity at 
1 kpc to be of the order of 600 km/s using the equation in \cite{2002ApJ...570..588R}, 
\begin{equation}
v_{esc}(r) = \sqrt{2} \; v_{circ} * [1 \;+ \; ln(1\;+ \;\frac{r_{max}}{r})]^{0.5}
\end{equation}
where $v_{circ} = \sqrt{2} \; \sigma_{*}$ is the circular velocity and 
$r_{max}$ = 100 kpc \citep{2011ApJ...732....9G}. This was calculated assuming a 
stellar dispersion $\sigma_{*}$ of 120 km/s for NGC 1365 \citep{2020A&A...634A.114C}. 
Thus, though the ionised gas is outflowing, it does not have the energy to escape 
the host galaxy’s potential. This is similar to the findings of other low and 
moderate luminosity Seyfert galaxies \citep{2014ApJ...792..101D,2019MNRAS.490.5860S}. 
The gas would just rain back down into the galaxy and help to relocate gas and 
dust and chemically enrich the galaxy. A caveat is that 
the outflow velocities considered here are the projected line of sight velocities 
and are not indicative of true outflow velocities which may be higher for high 
inclination angles of the outflow to the line of sight. However, \cite{2024ApJ...960...83S} find low metallicities in the central region of NGC 1365 which they attribute to a combination of inflow of metal poor gas and AGN feedback interrupting the star formation in the ring. Thus, though the AGN outflow may not escape the host galaxy it may have the potential to influence the star formation in the vicinity during its active phase.



\section{Summary}

The work presented here aims to understand the impact of AGN activity on star 
formation in NGC 1365 using new FUV and NUV data from UVIT combined with archival MUSE optical IFU data. The results of the work are summarised 
below
\begin{enumerate}
\item The UV data confirms existing literature studies of the SFR of the 
starburst ring in NGC 1365. We estimated a SFR of 2.68 M$_{\odot}$yr$^{-1}$ for 
the central 2 kpc from the FUV flux map corrected for extinction using the 
\cite{2000ApJ...533..682C} attenuation law assuming a R$_{\textup{v}}$ of 3.1. 
The estimated SFR surface densities of NGC 1365 are similar to other starbursts.
\item We found low FUV flux values (and hence low SFR) in the east and SE part of 
the star forming ring (see Fig. \ref{fig:figure-5} left panel). This 
finding based on the new UV observations reported here is in agreement with
those found from the very recent high resolution observations in the infrared and sub-mm
from JWST and ALMA. According to \cite{2023ApJ...944L..15S}, this in-homogeneity 
in SF characteristics on either side of the AGN is due to differences in the 
onset of star formation.
\item  The UV data revealed previously undetected diffuse UV emission 
co-spatial with the [OIII] 5007 \AA\; outflow of AGN origin 
(see Fig. \ref{fig:figure-4} top panel). Such a detection was possible only due 
to deep UV data combined with the high spatial resolution of UVIT. This points 
to a common origin for [OIII] 5007 \AA\ and diffuse UV emission in the northern and southern cones.
\item Ratios of UV fluxes to [OIII] continuum subtracted line fluxes indicate 
that the diffuse UV emission, which is co-spatial with the [OIII] 5007 \AA\; 
outflow, is of AGN origin (see Fig. \ref{fig:figure-4} bottom panel). This may 
be due to AGN light being scattered by electrons and dust particles in the bi-cone. 
\item We estimated the escape velocity at 1 kpc to be about 600 km/s. As the projected
velocity of the outflowing ionised gas is much lower than the escape velocity 
at 1 kpc, the gas does not have the energy to escape the host galaxy potential 
and will rain back into the galaxy.
\end{enumerate}
This work has shown the usefulness of UV in characterising the star formation
nature of AGN hosts. While UV data as a direct tracer of star formation has 
been used extensively in star formation studies, it has been less explored in 
the AGN domain. With the advent of JWST, rest frame UV studies of large 
statistical datasets will throw new light on the prevalence of AGN feedback and 
its impact on host galaxies.


\begin{acknowledgments}
\it{This publication uses data from the AstroSat mission of the Indian Space Research Organization (ISRO), archived at the Indian Space Science Data Centre (ISSDC). This publication uses UVIT data processed by the payload operations centre at IIA. The funding provided by the Humboldt Foundation, Germany is thankfully acknowledged. K.S.K. acknowledges the studentship at the European Southern Observatory, Garching.  TPA acknowledges the support of the National Natural Science Foundation of China (grant nos. 12222304, 12192220 and 12192221).}
\end{acknowledgments}

%

\vspace{5mm}
\facilities{AstroSat (UVIT), VLT(MUSE). UVIT was built in collaboration between IIA, IUCAA, TIFR, ISRO and CSA.}


\software{Astropy \citep{2013A&A...558A..33A,2018AJ....156..123A}, 
        MPDAF \citep{2016ascl.soft11003B},
          Cloudy 17.0 \citep{2017RMxAA..53..385F}
          }





\appendix
\section{[OIII] and H$\alpha$ flux maps}
\begin{figure*}[hbt!]
\centering
\vbox{
     \hbox{
          \includegraphics[scale=0.5]{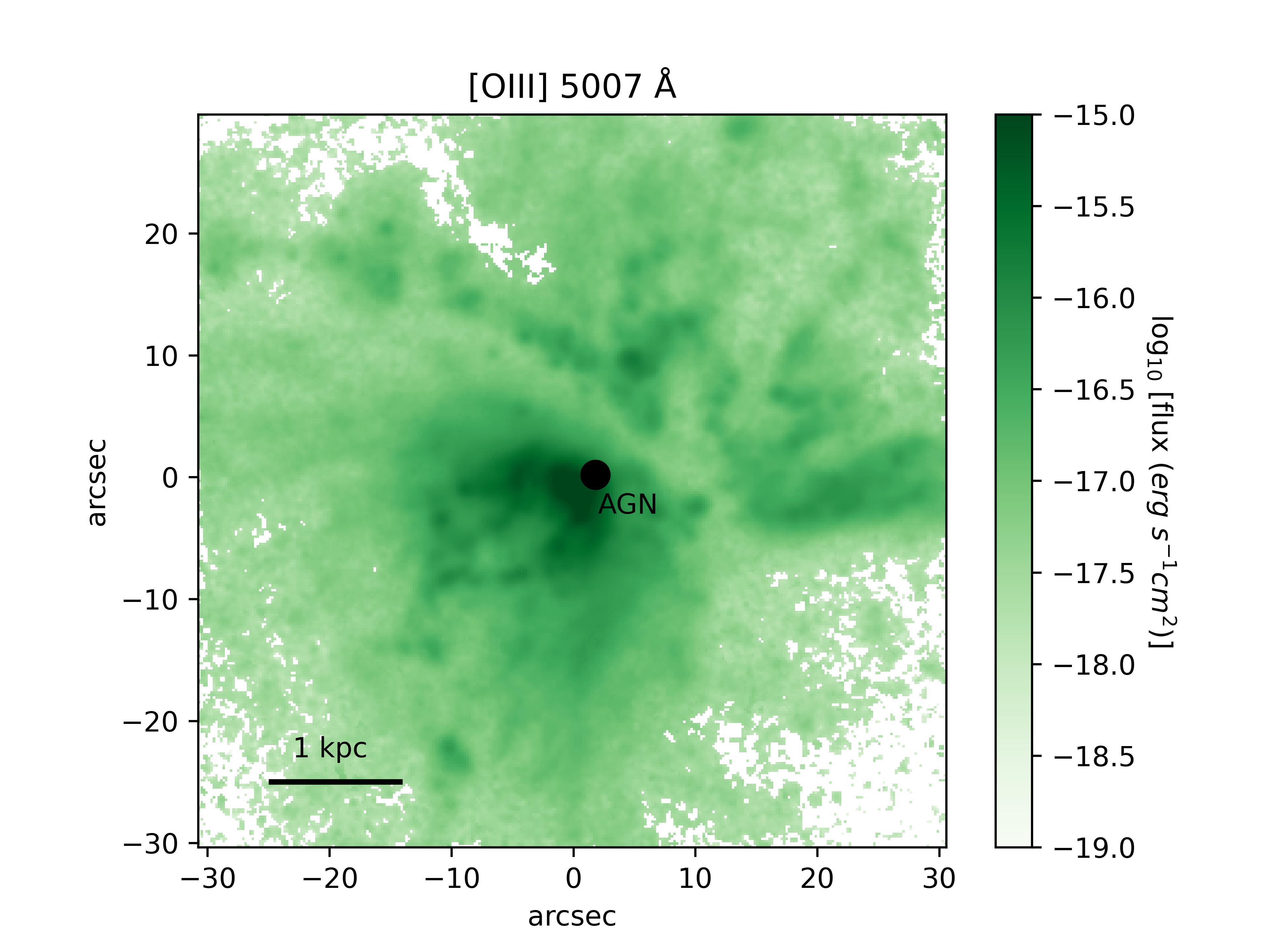}
          \includegraphics[scale=0.5]{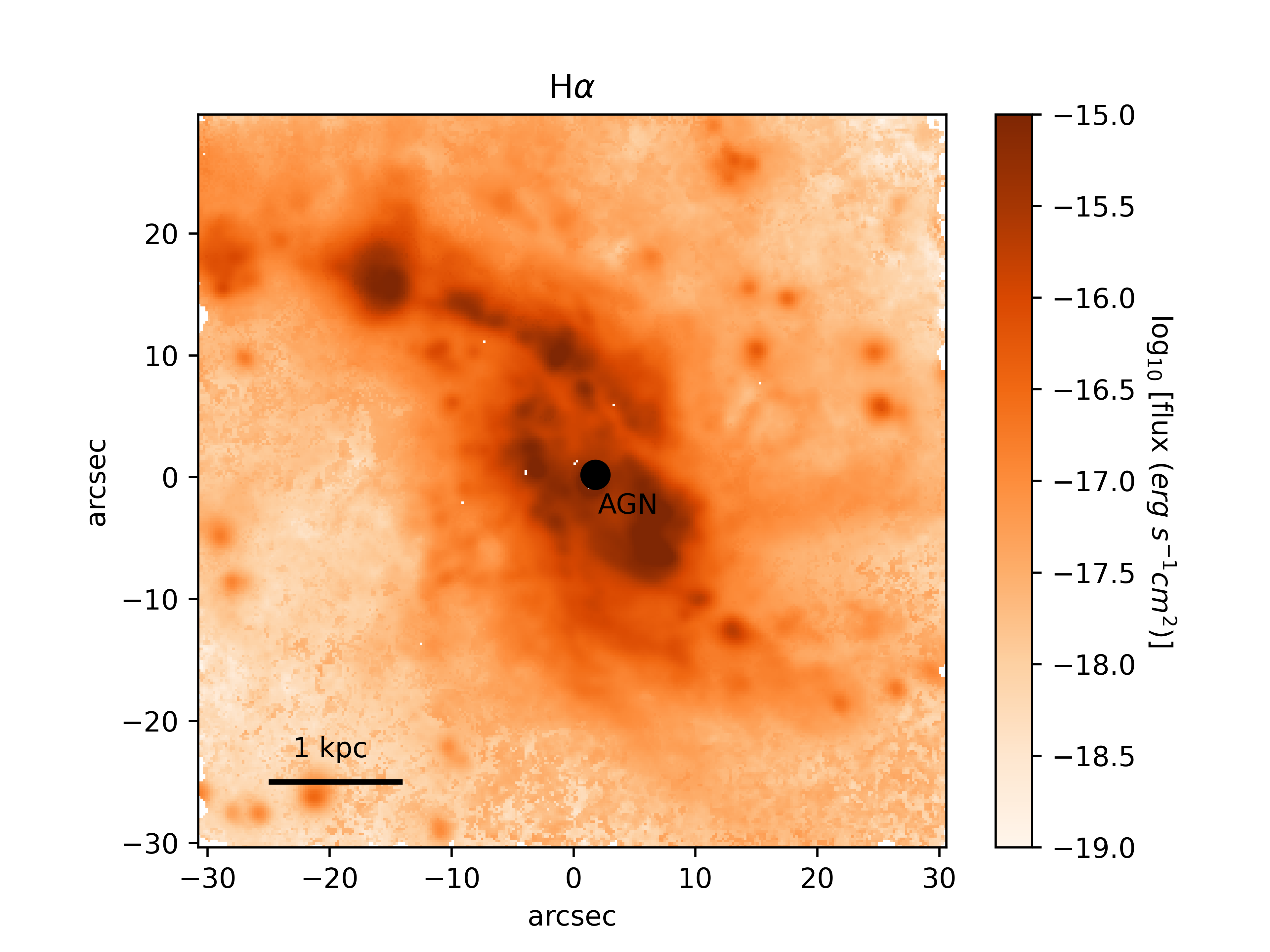}
          }
}
\caption{Flux maps of [OIII] and H$\alpha$ derived from the MUSE data}
\label{fig:figure-20}
\end{figure*}

Fig. \ref{fig:figure-20} shows the [OIII] and H$\alpha$ flux maps, we derived from the MUSE IFU data.

\section{Starburst ring and AGN double cone}
\begin{figure*}[hbt!]
\centering
\vbox{
     \hbox{
          \includegraphics[scale=0.45]{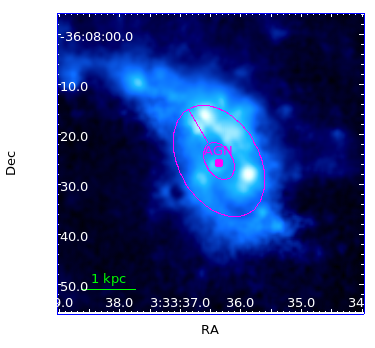}
          \includegraphics[scale=0.45]{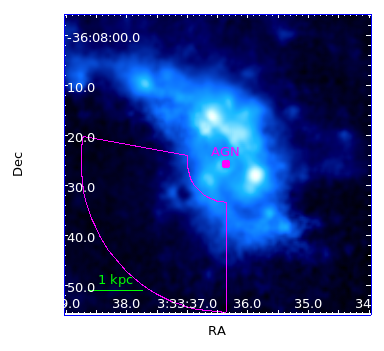}
          \includegraphics[scale=0.45]{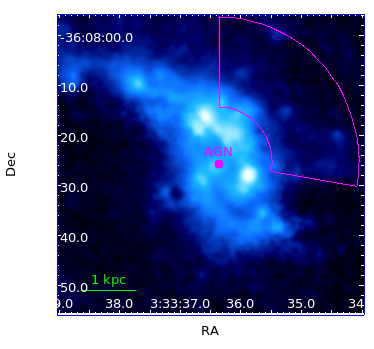}
          }
}
\caption{This figure shows the three regions used to determine the UV-optical flux ratios in Fig. \ref{fig:figure-4} overlaid on the intrinsic UVIT FUV flux map. The left panel shows the 12 arcsec starburst ring after masking the central AGN. This region is also used to determine the SFR of the ring in Section 4.2. The middle and right panels show the SE and NW cones respectively. }
\label{fig:figure-44}
\end{figure*}

Fig. \ref{fig:figure-44} shows the intrinsic FUV flux map with three regions of interest overlaid, from which the UV-optical flux ratios in Fig. \ref{fig:figure-4} have been determined.

\bibliography{ref_ngc1365}{}
\bibliographystyle{aasjournal}



\end{document}